\documentclass{ws-procs9x6}

\begin{document}
\newcommand{\dzero}     {D0}
\title{Results on top quark physics at D\O\ }

\author{Yvonne Peters}

\address{for the D\O\ Collaboration\\
University of Wuppertal \\
Gaussstrasse 20, 42097 Wuppertal \\
E-mail: peters@fnal.gov }

\maketitle

\abstracts{ In 1995, the top quark was discovered at the Fermilab Tevatron Collider by the CDF and D\O\ collaborations
in the $p\bar{p}$ collisions at $\sqrt(s)=1.8$~TeV using about $50\text{~pb}^{-1}$ of data per experiment. We present the studies of the
top quark properties based on $1\text{~fb}^{-1}$ of data collected by the D\O\ experiment at a center of mass energy of $1.96$~TeV.
The increased statistics and higher collision energy allow to perform precision measurements of the top quark production and decay characteristics and open the possibility to probe physics beyond the Standard Model in the top quark sector.
The presentation  mainly focuses on the measurement of the top pair production cross section and branching fraction of the top quark decays, a search for $t\bar{t}$ resonances, and the measurement of the top quark pair production cross section ratio and its interpretation in terms of new physics.
}

\section{Introduction}
In 1995, the CDF and D\O\ Collaborations discovered the top quark, the last missing piece in
the quark sector of the Standard Model (SM). The
top quark is the heaviest of all quarks and its mass of $172.6$~GeV is now known with a
precision of $0.8\%$~\cite{topmass}. 
Due to its short lifetime of about
$0.5\cdot 10^{-24}$~s~\cite{pdg} the top quark decays before
hadronization. This allows to measure the
 properties of a naked quark without any dilutions due to soft interactions. Several such measurements are presented in the following. 

\section{Top quark production and decay}
The top quark can be produced in pairs via the strong interaction or as a single
top quark via the electroweak interaction. Details about the latter can be found
in Ref.~\cite{stopPRL}.
At the Tevatron the top quark pairs
are produced through gluon fusion (15\%) 
and $q\bar{q}$ annihilation (85\%). In the SM top quarks decay almost
exclusively into a $W$ boson and a $b$-quark. 

Different final states of the $t\bar{t}$ pair are classified according
to the decay of the two $W$ bosons.  The classification separates events as
$\tau$+lepton (5\%), all hadronic (44\%), $\tau$+jets (15\%), lepton+jets (30\%) and
dileptonic (5\%). The lepton+jets channel consists of exactly one
isolated electron or muon, at least four jets and missing transverse energy to
account for the neutrino. The channel has a signal over background ratio of about
one and shows the best
combination of large statistics and clear signature. The main
background in this channel comes from $W$+jets events. 
Dileptonic events, consisting of two isolated leptons, at
least two jets and high missing transverse energy, are the purest events with
a signal over background of about three, but
suffer from low statistics. The main background in this channel originates
from $Z$+jets production.  

Since each $t\bar{t}$ event has at least two $b$-jets, $b$-jet identification
is a powerfull tool to distinguish $t\bar{t}$ signal from background~\cite{btagging}.  The $B$-hadron travels several
millimeters in the detector before it decays, producing a secondary
vertex displaced from the hard interaction vertex and tracks with high impact
parameters. Properties of both are used as an
input to the Neural Network $b$-tagging algorithm developed in \dzero. One can
reach a high performance, e. g. of  a $b$-tag efficiency of $54\%$ at
a fake rate of $1\%$.

\section{Top quark pair production cross section}
\begin{figure}[t]
\centerline{\epsfxsize=2.45in\epsfbox{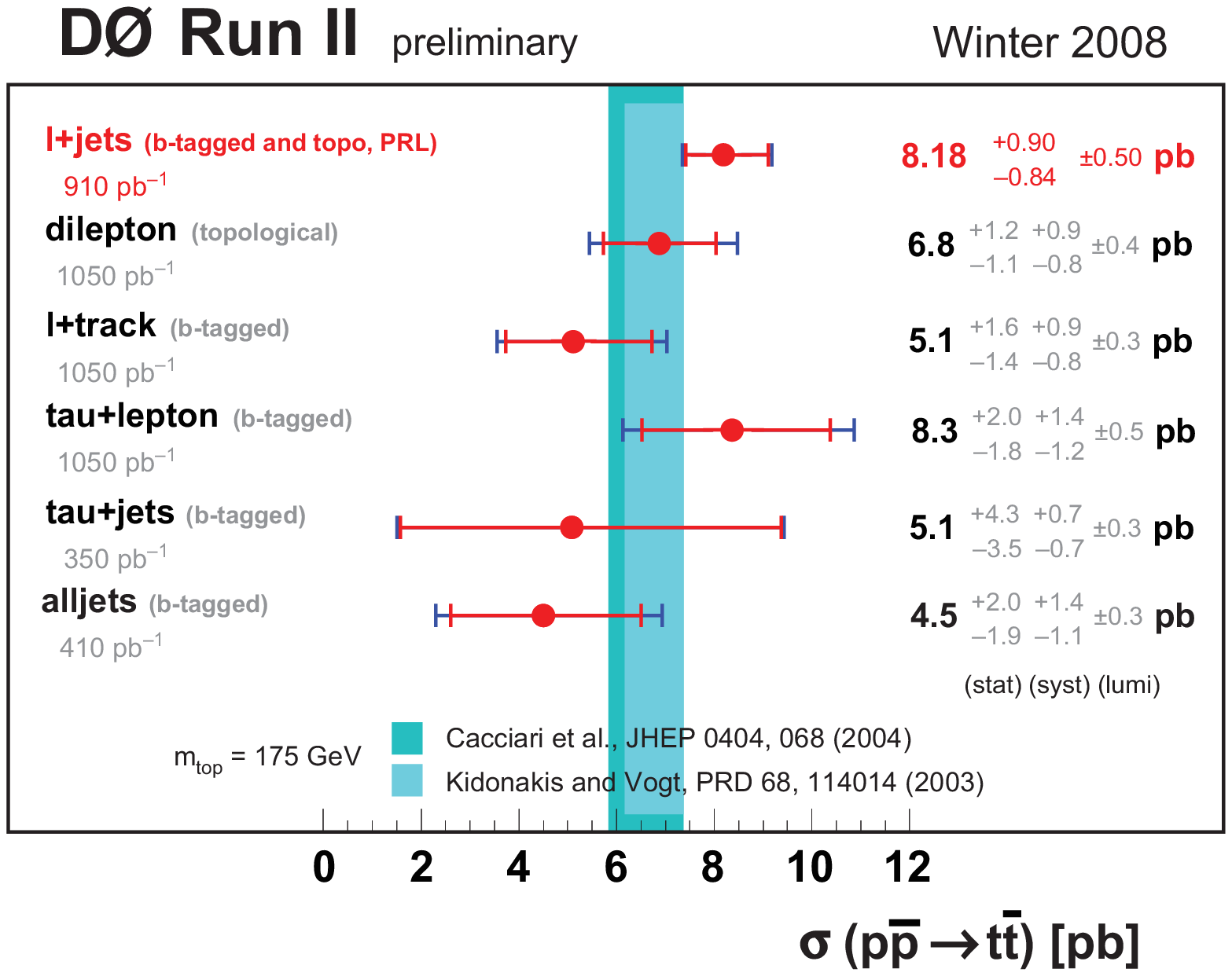}\epsfxsize=1.95in\epsfbox{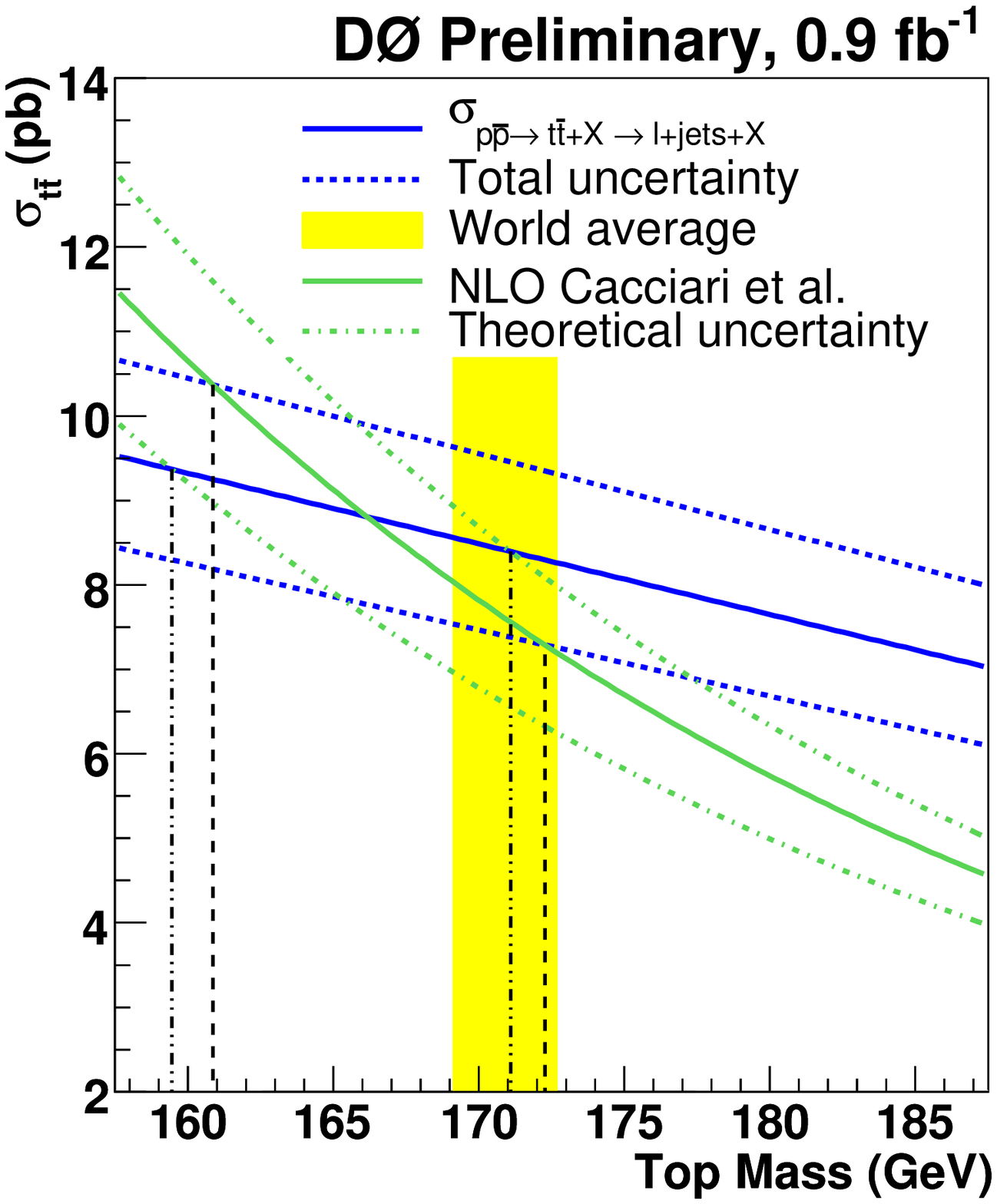}}   
\caption{Left: Measured $t\bar{t}$ cross sections in various decay
  channels. Right: Theoretical and measured cross section as function of top
  quark mass. \label{xsections_all}}
\end{figure}

The top quark pair production cross section has been measured in several decay channels, in
order to increase the statistics and to check their consistency. Processes
beyond the SM can lead to differences in the measured $t\bar{t}$ cross section
for different channels.
Figure~\ref{xsections_all} (left) shows a summary of the cross section
measurements with up to 1~$\text{fb}^{-1}$ of data. The
measurement in the lepton+jets channel with $b$-tagging is the most precise single cross section
measurement with $~11$\% relative uncertainty excluding the luminosity error. 
                           
All cross section 
measurements are comparable with each other and agree with the Next-to-Leading Order (NLO) SM expectation of $6.77\pm0.6$~pb
at a top mass of $175$~GeV~\cite{SMtheory}.

\section{Top quark mass from cross section}
Due to the high precision of the $t\bar{t}$ cross section measurements it is
worth to use it to extract the top mass~\cite{massfromxs}. The advantage
of this method over the standard top mass measurements is
its simplicity and a clear theoretical interpretation of the extracted mass
as the pole mass. The disadvantage is the larger uncertainty. We compare the theoretical
cross section calculation at NLO including higher order resummations, which shows a significant dependence on the top
 mass,  to the experimental cross section.  Figure~\ref{xsections_all} (right)
shows the  calculation from
Cacciari et al, and the measured cross section in the lepton+jets channel as a function of the top
mass. Their intersection point corresponds to a mass value of
$166.1^{+6.1}_{-5.3}\text{(stat+syst)}^{+4.9}_{-6.7}\text{(theory)}$~GeV.  
This value agrees with the world average.

\section{Top pair production through resonance}
\begin{figure}[t]
\centerline{\epsfxsize=2.20in\epsfbox{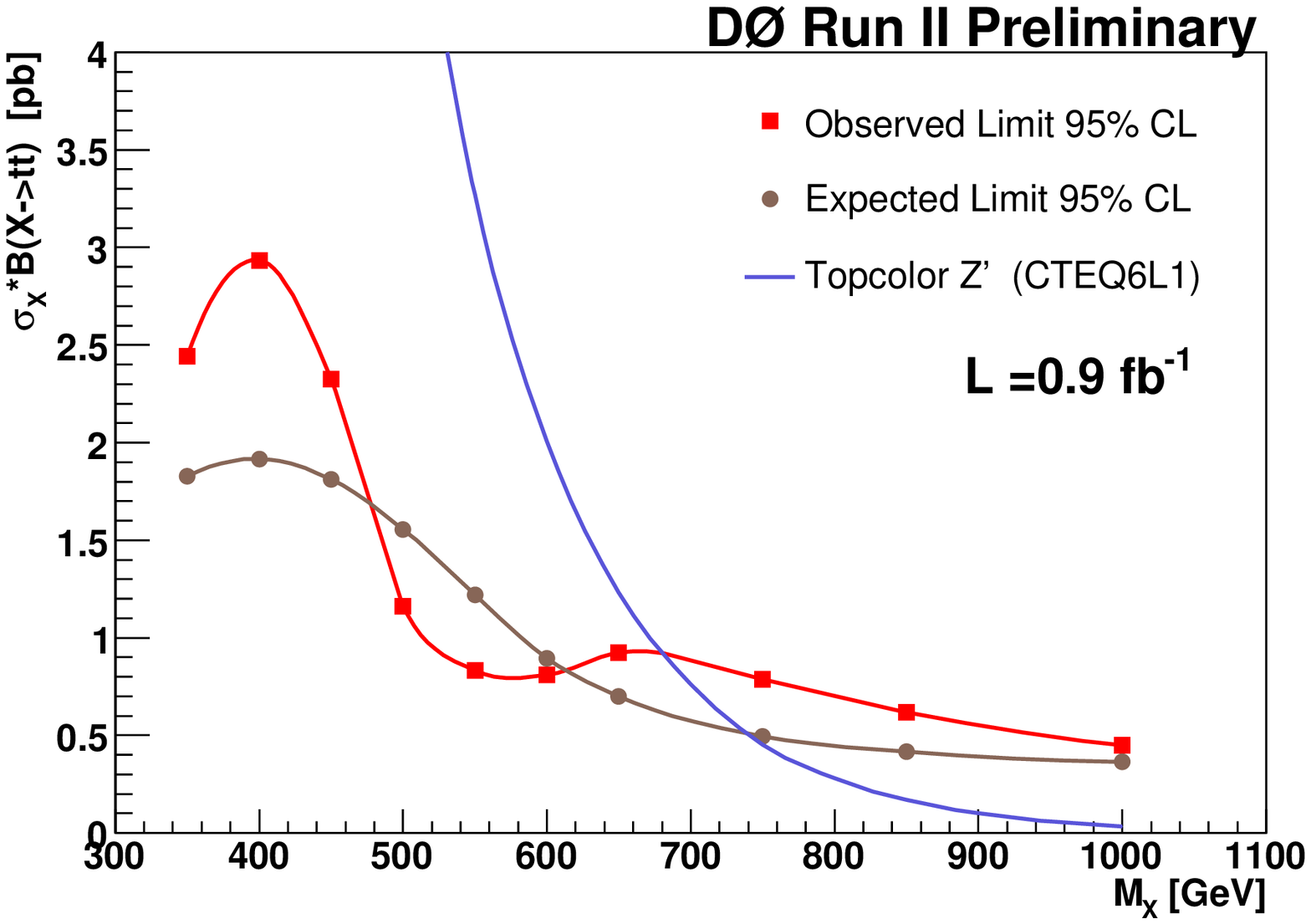}\epsfxsize=1.60in\epsfbox{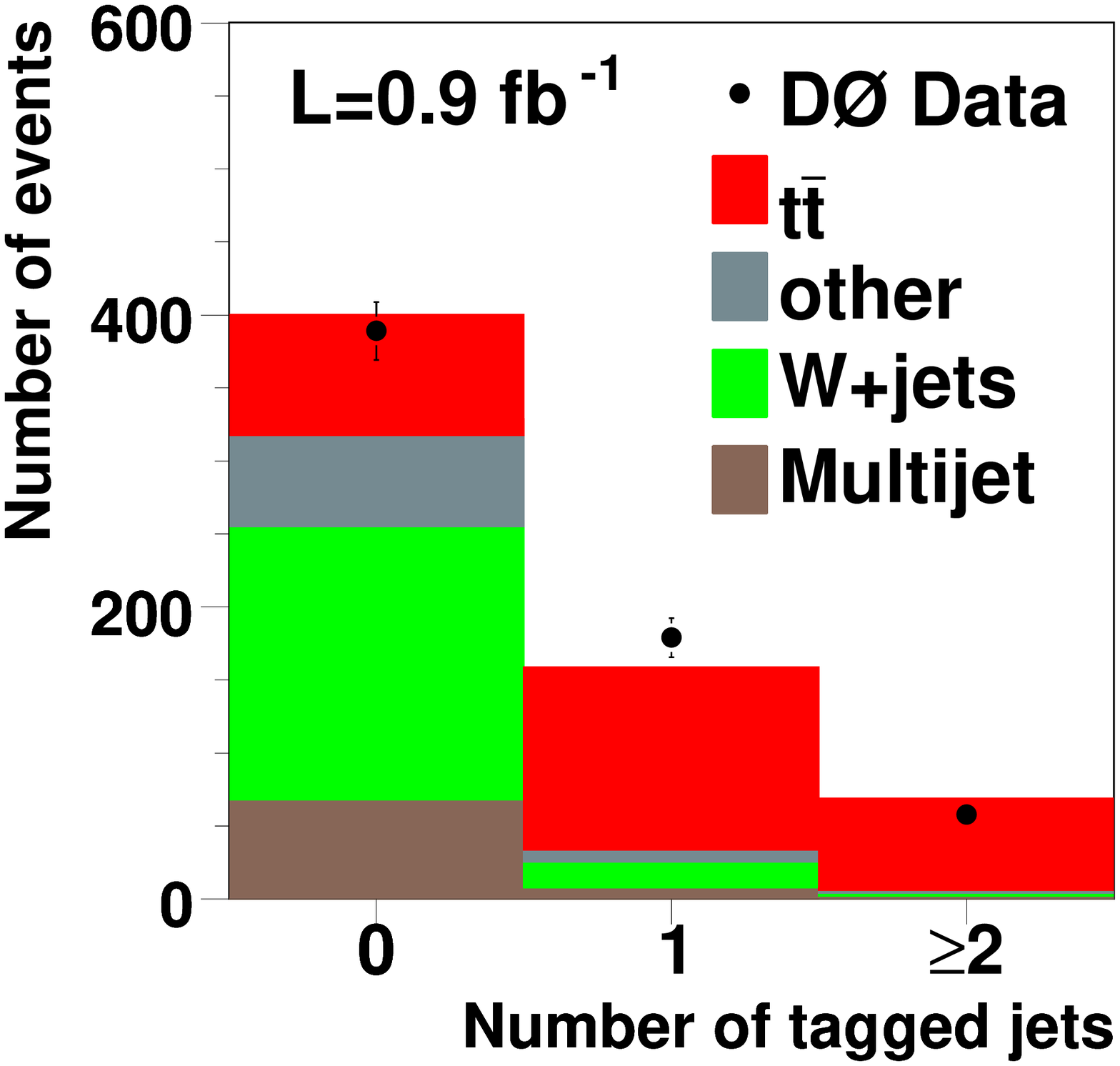}}   
\caption{Left: Observed and expected limits on  $\sigma_X\cdot B(X\rightarrow
t\bar{t})$ versus mass of $X$. Right: Expected and observed number of events
  versus number of $b$-tagged jets for measured $\sigma_{t\bar{t}}$ and $R$. \label{figure2}}
\end{figure}

In the SM no resonant $t\bar{t}$ production is predicted. A generic search
for narrow resonances $X$ decaying into $t\bar{t}$ was performed, by looking for bumps in the invariant mass
distribution $m_{t\bar{t}}$~\cite{ttres}. The event selection is based on the
lepton+jets channel, requiring at least four jets and at least one $b$-tag. 
As no enhancements in $m_{t\bar{t}}$ is observed, limits on the cross
section times branching ratio  $\sigma_X\cdot B(X\rightarrow
t\bar{t})$  as a function of the resonance mass $m_X$ are set. The expected
limit for the SM and the observed limit are shown in
Fig.~\ref{figure2} (left), together with the predicted dependence of $\sigma_X\cdot B(X\rightarrow
t\bar{t})$  on the $Z'$ boson mass in  top assisted technicolor models~\cite{technicolor}. The measured
 limits can be translated into a lower limit on
the mass of a $Z'$ boson  of $m_{Z'}>680\text{~GeV at 95\% C.L.}$ 

\section{Simultaneous measurement of $\sigma_{t\bar{t}}$ and $R$}
In the SM the ratio of branching fractions $R=\frac{B(t\rightarrow
  Wb)}{B(t\rightarrow Wq)}  = 
\frac{\mid V_{tb}\mid^2}{\mid V_{tb}\mid^2 + \mid V_{ts}\mid^2 + \mid
  V_{td}\mid^2} $ with $q$ being any down-type quark, is highly constraint by
  the unitarity of the CKM matrix and the requirement of $|V_{tb}|=0.999100^{
  +0.000034}_{ -0.000004}$~\cite{pdg}. 
 For the $t\bar{t}$
cross section measurement using $b$-jet identification $R$ is assumed to be one.
The simultaneous  measurement  of $\sigma_{t\bar{t}}$ and $R$ allows to probe
  both quantities without this assumption~\cite{Rmeas}. The measurement
  uses the fact that the number of identified
$b$-jets in $t\bar{t}$ events depends on $R$. For lepton+jets events with
  exactly three and at least four jets samples of no, one and at least two
  $b$-tagged jets are separated. In case of no identified $b$-jets in the event complementary information from a topological discriminant
is used, yielding
\begin{eqnarray*}
\sigma_{t\bar{t}} & = & 8.18^{+0.90}_{-0.84}\text{(stat+syst)}\pm0.5\text{(lumi)}\text{~pb}
\\
R &=& 0.97^{+0.09}_{-0.08}\text{(stat+syst).}
\end{eqnarray*}  
Figure~\ref{figure2} (right) shows the expected and observed number of events
versus the number of $b$-tagged jets for the measured value of $\sigma_{t\bar{t}}$ and
$R$. 
As $R$ is compatible with the SM we set limits on $R$ and $|V_{tb}|$ of
$R>0.79$ at 95\%C.L. and $|V_{tb}|>0.89$ at 95\%C.L. 

\section{Top pair production cross section ratio}
All results presented so far use the assumption that the top quark always
decays into a $W$ boson $B(t\rightarrow bW)=100\%$.  
The ratio of cross sections measured in the lepton+jets and dilepton channels 
$R_{\sigma}  =  \frac{\sigma(p\bar{p}\rightarrow
  t\bar{t})_{l+jets}}{\sigma(p\bar{p}\rightarrow  t\bar{t})_{dilepton}} $
 can be used to
  explore alternative options available beyond the SM~\cite{hplus}. Any
  deviation from $R_{\sigma}=1$ would indicate new physics, as e. g. the
  existence of a non-vanishing branching ratio  $B(t\rightarrow bX)$ with $X$ being any  particle but the $W$
  boson. In this analysis, $X$ is interpreted as a charged Higgs boson
  $H^{\pm}$, with mass close to the $W$ mass, $B(H^{\pm}\rightarrow cs)$ of 100\%
  and event kinematics similar to $t\rightarrow   bW$ decay. We derive
  $R_{\sigma}=1.21^{+0.27}_{-0.26}\text{(stat+syst)}$ using the 
  cross section measurement in the lepton+jets channel requiring at least four jets and  at least
  one $b$-tagged jet and the combined cross section  in the
  dileptonic final state of 
\begin{eqnarray*}
 \sigma(p\bar{p}\rightarrow t\bar{t})_{l+jets} & = &
  8.27^{+0.96}_{-0.95}\text{(stat+syst)}\pm 0.51\text{(lumi) pb \ \ \    and} \\  
\sigma(p\bar{p}\rightarrow t\bar{t})_{dilepton} & = &
 6.8^{+1.2}_{-1.1}\text{(stat)}^{+0.9}_{-0.8}\text{(syst)}\pm 0.4\text{(lumi)
 pb}.
\end{eqnarray*}
 In our model  $t\rightarrow   bH^{\pm}$ decays with  $B(t\rightarrow
  bH^{\pm})>0.35$ are excluded at 95\%
  C.L. 

\section{Summary}
New results on measurements and searches in the top quark sector with
$1\text{~fb}^{-1}$ of integrated luminosity have been presented. Some of the measurements are already
systematically limited, e. g. the top pair production cross section in the
lepton+jets channel. The high precision makes the top quark sector interesting
to search for new physics. In all searches performed so far no evidence for
new physics was observed. More searches and measurements can be found
at~\cite{ourresults}. With more than $3\text{~fb}^{-1}$ of integrated
luminosity on tape and more data to come and improved analysis methods we expect
many new precision measurements of top properties and searches in the top
sector in the near future.

\end{document}